\documentclass[]{spie} 
\usepackage{latexsym}
\usepackage{times,mathptm} 
\usepackage{graphicx} 
\usepackage{multicol}
\usepackage{rotating} 
\usepackage[verbose]{wrapfig}
\usepackage{epsfig}
\graphicspath{{/home/giulia/curioni/SPIE2002/papers/bkgd_flux/figures/}}
\newcommand{\g}{$\gamma$}

\newcommand{\po}{$^{40}$K}

\newcommand{\bc}{\begin{center}}
\newcommand{\ec}{\end{center}}
\newcommand{\be}{\begin{equation}}
\newcommand{\ee}{\end{equation}}
\newcommand{\bfg}{\begin{figure}}
\newcommand{\efg}{\end{figure}}
\newcommand{\bi}{\begin{itemize}}
\newcommand{\ei}{\end{itemize}}
\newcommand{\bt}{\begin{table}}
\newcommand{\enta}{\end{table}}
\newcommand{\keV}{\mbox{ke\hspace{-0.1em}V}}
\newcommand{\MeV}{\mbox{Me\hspace{-0.1em}V}}

\renewcommand{\deg}{\ensuremath{^\circ}}

\title{On the Background Rate in the LXeGRIT Instrument during the 2000 Balloon
Flight}  

\author{A.~Curioni\supit{a}, E.~Aprile\supit{a}, K.L.~Giboni\supit{a} ,
	M.~Kobayashi\supit{a}, \\ 
        U.G.~Oberlack\supit{b}, E.L.~Chupp\supit{c}, P.P.~Dunphy\supit{c},
	T.~Doke\supit{d},		 
        J.~Kikuchi\supit{d}, S.~Ventura\supit{e},  
    	\skiplinehalf
        \supit{a}Columbia Astrophysics Laboratory, Columbia University \\
        \supit{b}Rice University \\
        \supit{c}University of New Hampshire \\
        \supit{d}Waseda University, Japan \\
	\supit{e}INFN and Universit\`a di Padova, Italy
} 

\authorinfo{Further author information: (Send correspondence to: Alessandro
            Curioni) \\Columbia Astrophysics Laboratory, Columbia University,
            New York, NY, USA \\ 
            E-mail: curioni@astro.columbia.edu \hfill}

\begin{document} 
  \maketitle

\begin{abstract}
LXeGRIT is the first prototype of a novel Compton telescope for MeV \g-ray
astrophysics based on a Liquid Xenon Time Projection Chamber (LXeTPC), sensitive
in the energy band of 0.15 -- 10~\MeV. In this
homogeneous, 3D position sensitive detector, \g-rays with at least two
interactions in the sensitive volume of 2800 cm$^{3}$, are imaged as in a
standard Compton telescope.  \g-rays with a single interaction cannot be
imaged and constitute a background which can be easily identified and rejected.
Charged particles and localized $\beta$-particles background is also easily
suppressed based on the TPC localization capability with millimeter resolution.
A measurement of the total \g-ray background rate in near space conditions
and the background rejection power of the LXeTPC was a primary goal of the
LXeGRIT balloon flight program. We present here a preliminary analysis
addressing this question, based on balloon flight data acquired during the Oct
4-5, 2000 LXeGRIT balloon flight from Ft. Sumner, NM.  In this long duration (27
hr) balloon experiment, the LXeGRIT TPC was not surrounded by any \g-ray or
charged particle shield.  Single site events and charged particles were mostly
rejected on-line at the first and second trigger level.  The remaining count
rate of single-site \g-ray events, at an average atmospheric depth of
3.2~g~cm$^{-2}$, is consistent with that expected from atmospheric and diffuse
\g-ray background, taking into account the instrument mass model and response.
\end{abstract}


\section{LXeGRIT 2000 Balloon Flight} \label{sec:flight}

The Liquid Xenon Gamma Ray Imaging Telescope (LXeGRIT) uses a Liquid Xenon Time
Projection Chamber (LXeTPC) to image \MeV\ \g-rays from the energy deposits
and 3D spatial coordinates, measured on an event-by-event basis. The instrument
and results from laboratory and previous balloon flight experiments in 1997 and
1999 are discussed in various references
~\cite{EAprile.98.electronics,EAprile.2000.pisa,EAprile.2000a.SPIE,EAprile.2000.cgro,EAprile.2000b.SPIE}.  
Here we focus on an analysis of data acquired during the last LXeGRIT balloon
flight, when the  LXeTPC was used without any \g-ray or charged particle
shield. The flight took place on Oct 4--5, 2000, from Ft. Sumner, NM and lasted
27 hours, including ascent. Fig.~\ref{f:altitude} shows the altitude of the
balloon payload and the corresponding atmospheric depth, which varies between
5.7 and 3.2~g~cm$^{-2}$. 

\bfg[htb]
\centering
\psfig{file=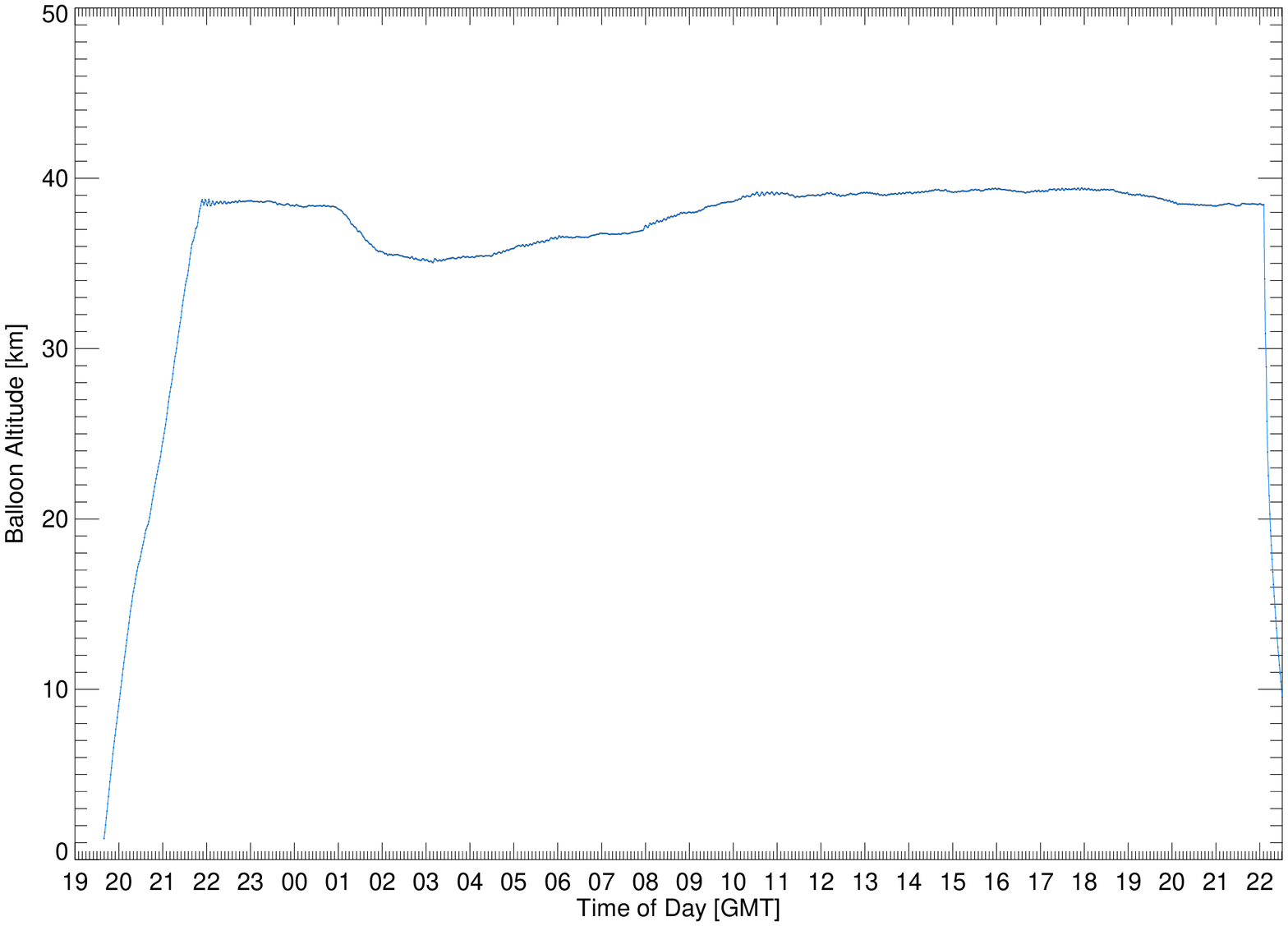,width=.4\linewidth,angle=90,clip=}
\\[1ex]
\psfig{file=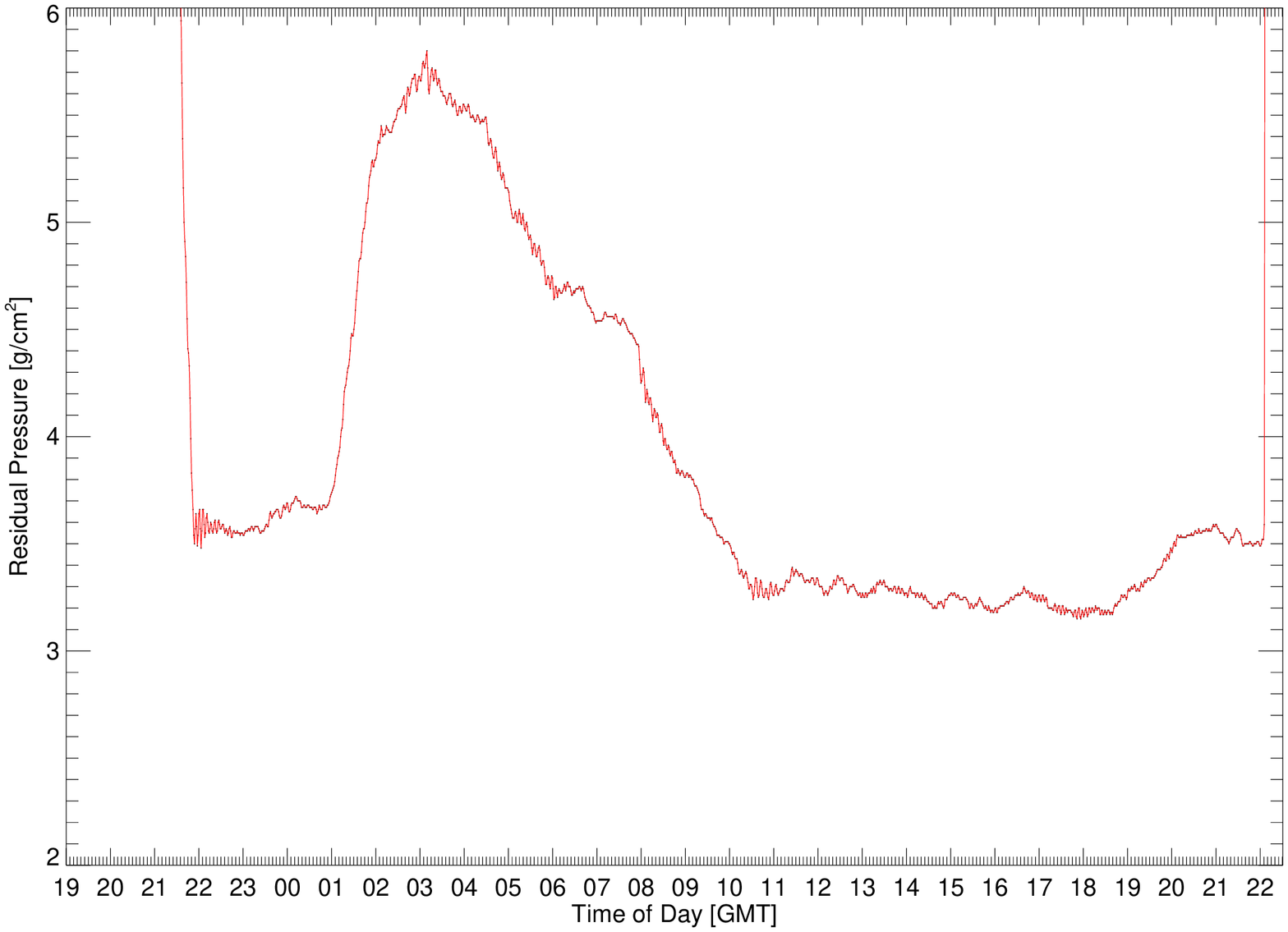,width=.4\linewidth,angle=90,clip=} 
\caption{\label{f:altitude} \emph{Top:} Altitude of the balloon payload during
the 2000 flight. \emph{Bottom:} Atmospheric depth.}
\efg
%

The zenith distance of known celestial \g-ray sources during flight is shown in
Fig.~\ref{f:sources_FOV}. The Crab nebula, among them the brightest in \g-rays,
was in the LXeGRIT field of view - which we define within an angular distance of
60\deg\ -  for more than 8 hours, from 7:00 to 15:30 UT.
\bfg[htb]
\centering
\psfig{file=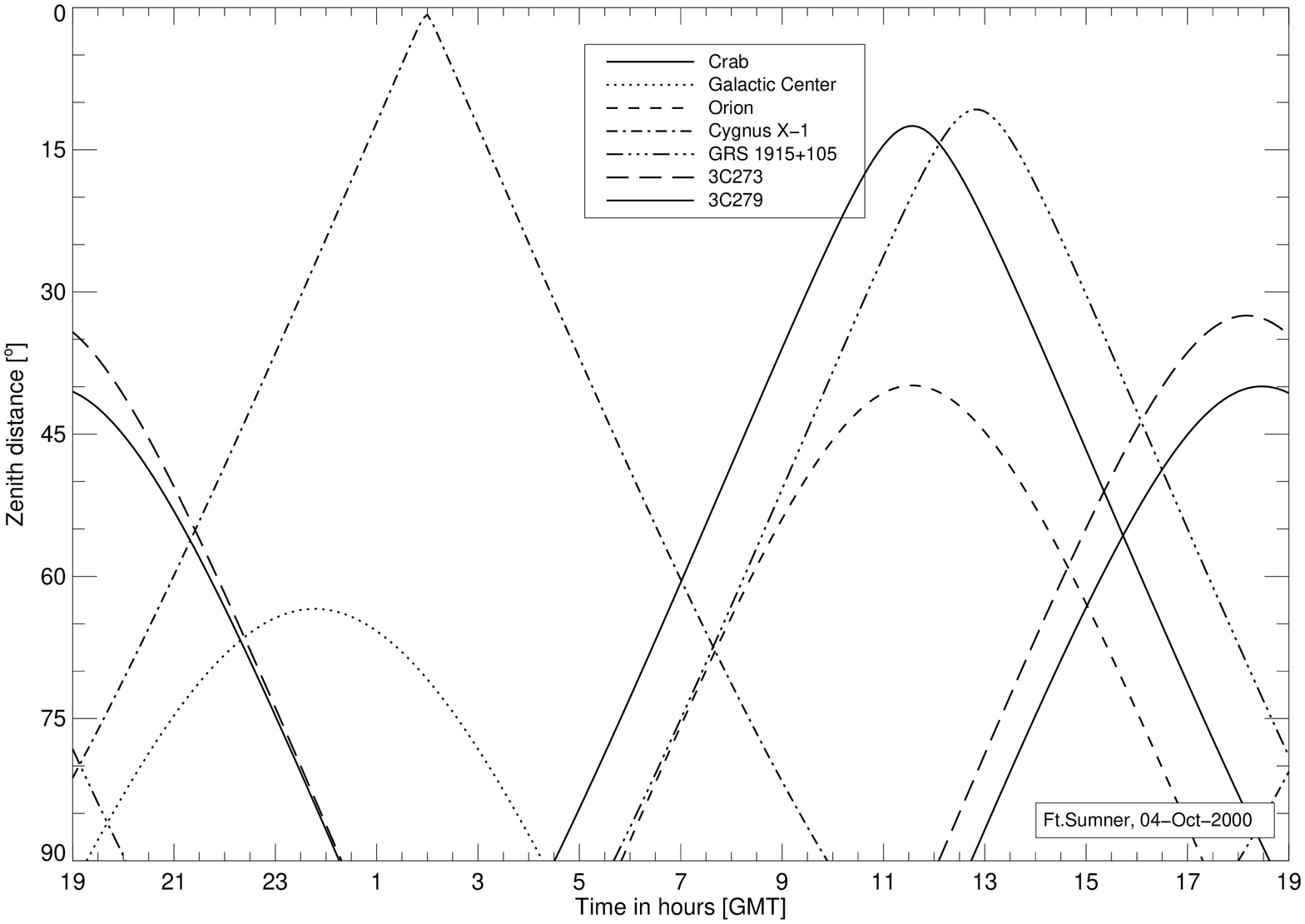,width=.4\linewidth,angle=90,clip=} 
\caption{\label{f:sources_FOV} Known celestial \g-ray sources in the LXeGRIT
field of view during the 2000 flight. }
\efg
The analysis presented in this paper is based on single-site events recorded
over 5 hours, from 12:00 to 17:00 UT, on October 5, 2000. In this period LXeGRIT
was at a stable altitude of $\sim$39~km or $\sim$3.2~g~cm$^{-2}$\ atmospheric
depth.  
 
The overall goal of this analysis is to understand the measured background count
rate in the LXeTPC. In  the 2000 flight configuration, the instrument was
sensitive in the energy range from $\sim$150~\keV\ to 10~\MeV.  
The LXePTC trigger rate, provided by the primary scintillation light signal, was
$\sim$600~Hz (Fig.~\ref{f:trg_rates}, top panel). This rate was nearly constant
throughout the flight, after ascent had been completed.
The livetime fraction of the on-board data acquisition system (DAQ), shown in
Fig.~\ref{f:trg_rates}, bottom panel, was $\sim$50~$\%$.  
Since the LXeTPC itself has a deadtime of about 50~$\mu$s~/~event and for a
rate of 600~Hz the deadtime fraction amounts to less than 5~$\%$, the instrument
deadtime is largely determined by the DAQ.
The DAQ processor was able to handle $\sim$300~Hz (Built and Rejected Event
rates are shown in Fig.~\ref{f:trg_rates_2}) - the rejection rate at the second
level trigger was $\sim$250~Hz, the rate of 
selected events $\sim$50~Hz, all of
them transmitted to ground or written to the on-board hard disks. 
The ``gaps'' visible in the rate vs. time plots correspond to the TPC cooling
periods during which the DAQ is turned off due to the increased noise level on
the anodes, while the PMTs remain operational. 
More details about on-line event rejection are given in Sec.~\ref{sec:trigger}.
\bfg[htb]
\centering
\psfig{file=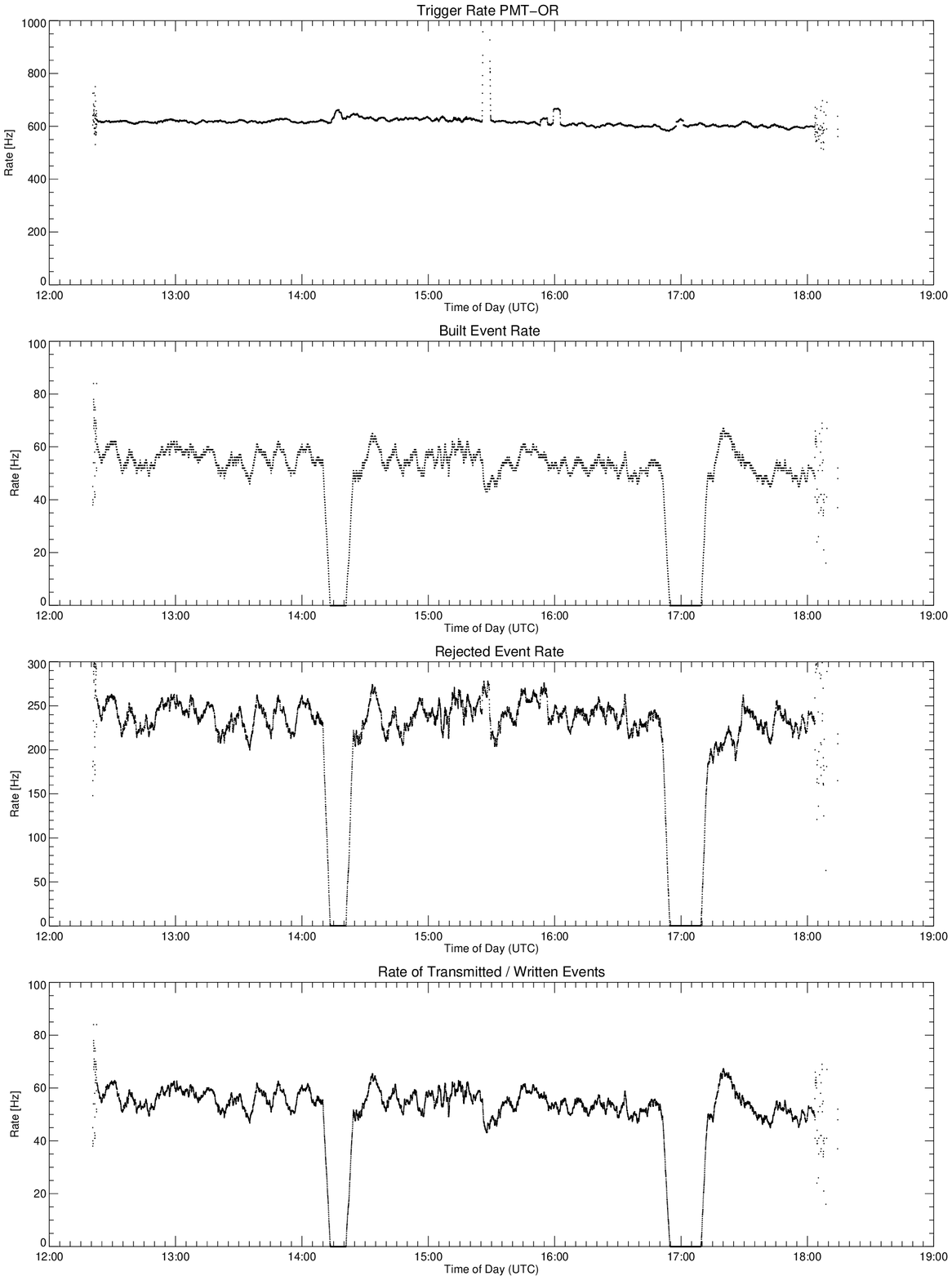,bbllx=60,bblly=565,bburx=566,bbury=732,width=.9\linewidth,clip=} 
\psfig{file=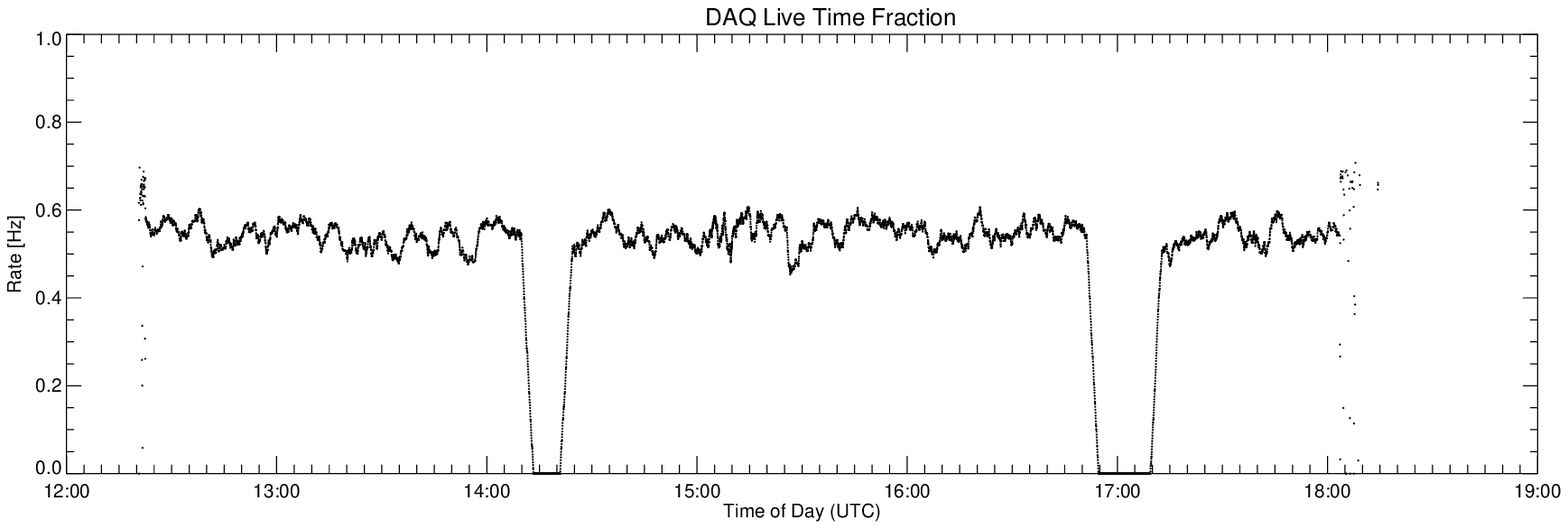,bbllx=60,bblly=565,bburx=566,bbury=732,width=.9\linewidth,clip=} 
\caption{\label{f:trg_rates} \emph{Top:} PMT-OR rate, \emph{bottom:} DAQ
livetime fraction.} 
\efg
\bfg[htb]
\centering
\psfig{file=SPIE02.5.ps,bbllx=60,bblly=224,bburx=566,bbury=560,width=.9\linewidth,clip=} 
\caption{\label{f:trg_rates_2} \emph{Top:} selected event rate, 
                               \emph{bottom:} rejected event rate.} 
\efg

\subsection{Event trigger and data acquisition} \label{sec:trigger}

The LXeGRIT event trigger works on two different levels: the \emph{first level
trigger} requires a signal from at least one of the four PMTs detecting the
light from the LXeTPC volume, while the \emph{second level trigger} selects
events based on the anodes~/~wires ionization signals recorded for each
triggered event. 
Decisions at the second trigger level are made necessary by the limited DAQ
speed (for a thorough and up-to-date account of the LXeGRIT DAQ see
Aprile~et~al.~\cite{EAprile.2001.IEEE}) which imposes an upper limit to 
the detector livetime, i.e. the maximum rate of events transmitted through
telemetry or written to the on-board disk (about 150 evts~sec$^{-1}$, depending
on the event size). \\   

The \emph{first level trigger} is provided by the OR of the four PMTs. This fast
light signal starts the DAQ and marks the beginning of the drift time
measurement. To avoid pile-up of independent events, it is required that 
there is no PMT signal within 50~$\mu$s before the event.
This trigger level allows a fast decision but it is rather insensitive to
different event topologies or energies. Ideally a high trigger efficiency at
this level would be desirable but in practice, given the soft energy spectrum of
the atmospheric and cosmic diffuse \g-ray background this would be equivalent to
accept a dominant fraction of low energy (150-500~\keV) events which are of very
little use for a Compton telescope like LXeGRIT with an energy threshold of
$\sim$150~\keV.   
The light trigger efficiency (LTE) for the 2000 flight has been measured for
energies up to 2~\MeV\ and spatially resolved with a few millimeter
granularity. In flight configuration the LTE averaged over the active volume 
was 10-20~$\%$ in the energy range 1-2~\MeV, steeply decreasing at lower
energies. 
A similar measurement has been described in
Oberlack~et~al.~\cite{UOberlack.2001.IEEE} for energies up to 500~\keV\ and for
the 1999 LXeGRIT settings of the light readout. 
The LTE is quite dependent on the location of the interaction inside the
fiducial volume because of the varying solid angle viewed by the PMTs.
Fig.~\ref{f:lightz} shows the solid angle effect on
light collection by the z-dependence measured for one of the four PMT's, at an
energy of 1.836~MeV. The strong dependence on interaction location requires a 3D
spatial $times$ 1D energy mapping of the efficiency,

The solid angle effect on light collection is shown by the z-dependence
measured for one of the four PMTs (Fig.~\ref{f:lightz}).
The strong dependence on interaction location makes necessary a 4-D
(x, y, z, energy) map of the light trigger efficiency, now fully implemented in
the Monte Carlo simulation. 
\bfg[h]
\centering
\psfig{file=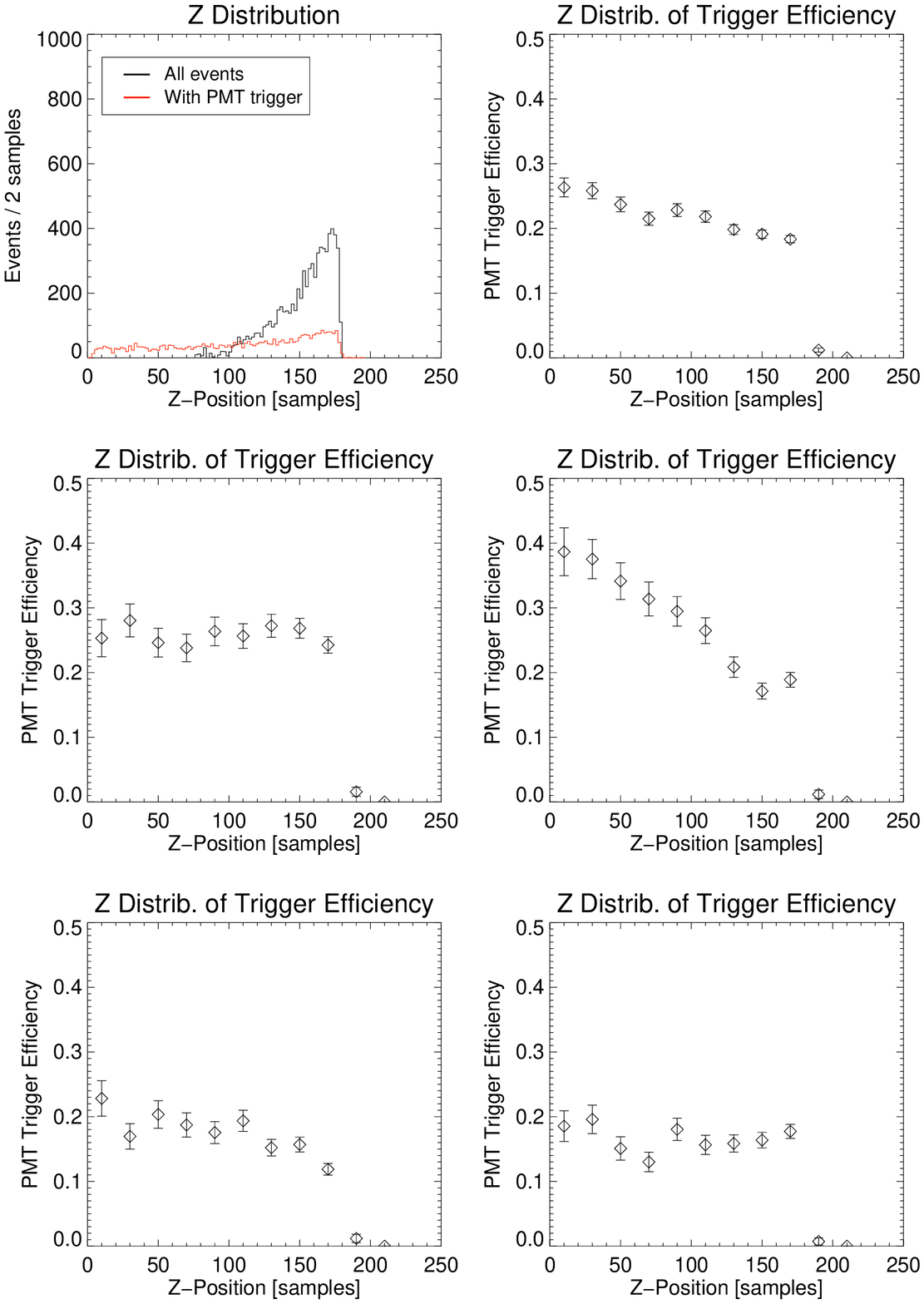,bbllx=320,bblly=280,bburx=550,bbury=483,width=.4\linewidth,clip=} 
\caption{\label{f:lightz} LTE vs. z position. The PMT is located below z=0 and
the solid angle effect is clearly visible.}
\efg

The \emph{second level trigger} performs the following on-line selections: 
a. it requires a minimum and a maximum number of wire hits on each view (x and
y) -- with this MINMAX selection, the minimum rejects mostly single site events,
the maximum rejects extended tracks due to high energy charged particle crossing
the active volume; 
b. it requires a minimum signal amplitude above baseline on at least one anode -
this requirement mainly rejects noise; 
c. it requires that the anode signal does not saturate the FADC, i.e. energy
deposit on a single anode less than $\sim$10~\MeV\ (flight 2000 settings) -
this mainly rejects charged particles which deposit a large amount of energy in
the active volume.
\bfg[htb]
\centering
\psfig{file=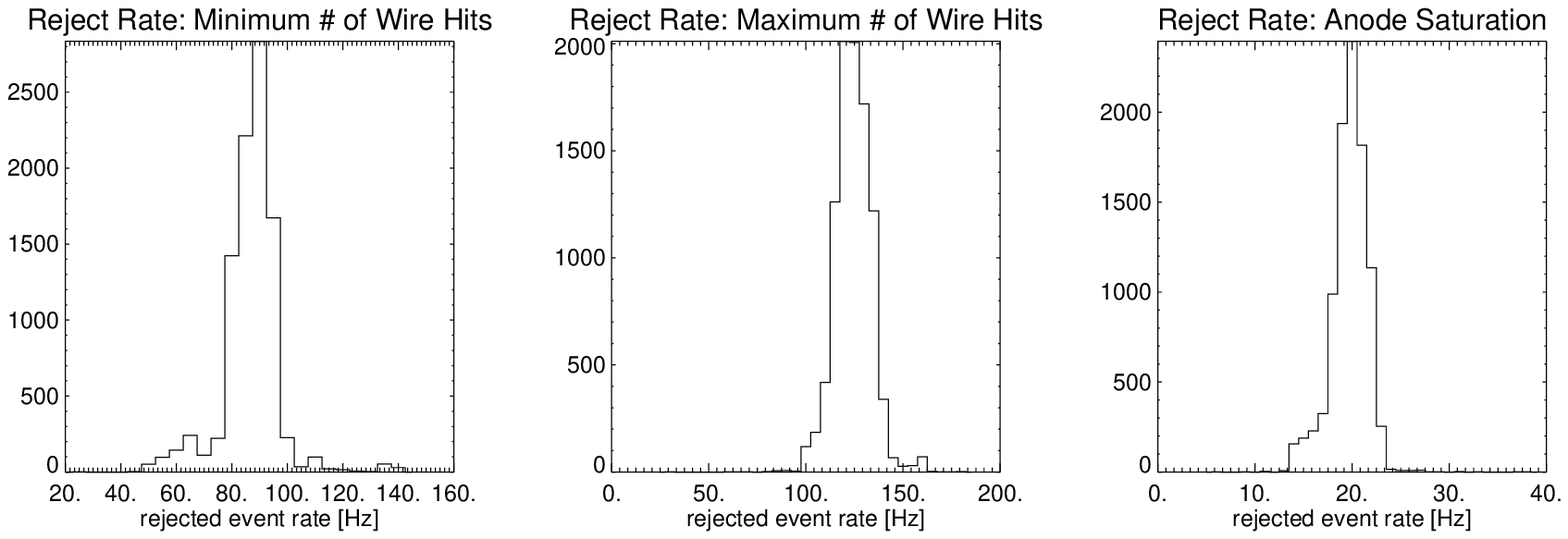,bbllx=73,bblly=561,bburx=561,bbury=731,width=.9\linewidth,clip=} 
\caption{\label{f:rejects} Distributions of rejected event rates due to specific
selection at the second level trigger. \emph{Left:} minimum number of wire hits
both in x and y. \emph{Center:} maximum number of wire hits in x or y. {\it
Right:} anode saturation, i.e. an energy deposit larger than 10~\MeV.} 
\efg
Fig.~\ref{f:rejects} shows the impact of specific on-line selections on the
flight data. 

These on-line selections determine the type of event topologies accepted by
LXeGRIT for further analysis. This option is relevant for a Compton telescope,
since Compton imaging requires at least two interactions (Compton scattering
followed by photoabsorption). 
In LXeGRIT, events with a single energy loss localized in one site of the
sensitive volume (single site events) are easily recognized and rejected as
background for source imaging. These events include either low energy \g-rays
which are photoabsorbed or higher energy \g-rays which interact only once before
escaping, as well as $\alpha$- or $\beta$-particles from internal background.
The rejection power of the on-line MINMAX selection for single site events as a
function of energy is shown in Fig.~\ref{f:minmax}. It is better than 60~$\%$
for energies up to 4~\MeV, i.e. for a large majority of the single site events
(see Fig.~\ref{f:intbkgd_2}). While effective, this second level trigger
selection has the major drawback of taking DAQ time.
Any remaining single site event can be rejected in the off-line analysis with
efficiency very close to 100~$\%$.

\bfg[h]
\centering
\psfig{file=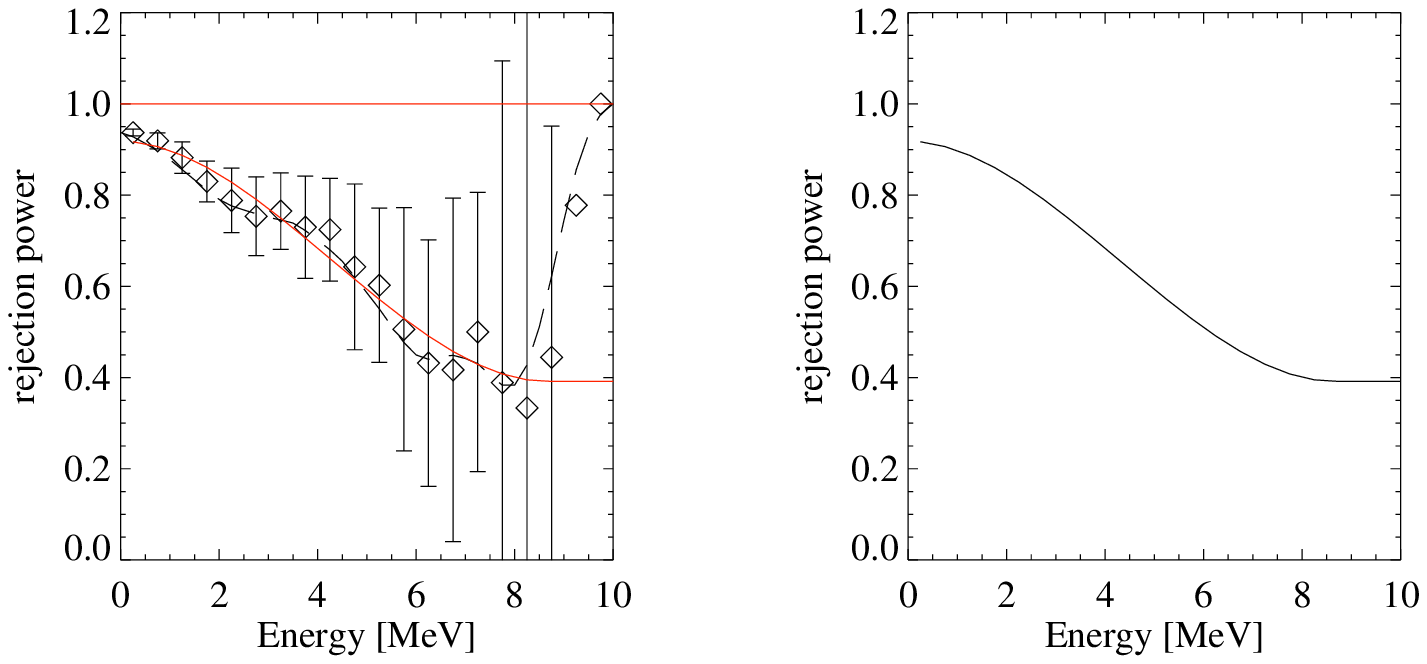,bbllx=317,bblly=526,bburx=500,bbury=716,width=.35\linewidth,clip=} 
\caption{\label{f:minmax} Rejection power of the MINMAX criterion for single
site events.}
\efg


\subsection{Off--line data reduction} \label{sec:offline}

A \emph{raw event} consists of the digitized (62-x and 62-y) wire and 4 anode
signals, sampled at a frequency of 5~MHz. Going from the raw data to a data
structure providing energy and 3D-position for each interaction, i.e. (x$_1$,
y$_1$, z$_1$, E$_1$), ..., (x$_n$, y$_n$, z$_n$, E$_n$), where $n$ is the
interaction multiplicity, requires several steps of data processing. \\
%
In the first place, the analysis program identifies and rejects noisy events,
events with interactions outside the fiducial volume, non-\g\ events in general
(e.g. charged particles).\\  
A large fraction of the flight data are rejected at this stage of the analysis;
in the end, about 20~$\%$ of the events are selected. Selected events are sorted
according to their interaction multiplicity - this classification comes quite
natural for a Compton telescope like LXeGRIT, since at least two interactions
are required for Compton imaging and at least three interactions are required to
determine the unique time sequence of the event. Since the Compton imaging
capability of LXeGRIT is not exploited in the measurement presented here, we
omit a discussion of the various issues pertaining to LXeGRIT as a Compton
telescope, for which we refer to Oberlack~et~al.~\cite{UOberlack.2000.SPIE}.\\
Having (x$_i$, y$_i$, z$_i$) for each interaction, fiducial volume selections
can be applied. This kind of selection is very powerful (rejection power
$\sim$1) in rejecting charged particle entering the TPC and localized (non-\g)
internal background (see Sec.~\ref{sec:infl_espec} and Sec.~\ref{sec:comp}). 
A full imaging detector provides in fact a unique signature for charged
particles. Since they lose energy continuously ionizing the active medium,
the point where they enter the active volume is easily identified.
The fine granularity allows a modest loss in fiducial volume. In this analysis,
events with at least one interaction on the two outer wires (x or y coordinate -
two wires correspond to 6~mm) or in the highest or lowest 3~mm along the z
coordinate are rejected. After selections the fiducial volume is reduced to
18.8$\times$ 18.8$\times$ 6.4~cm$^3$ (88~$\%$ of the original fiducial
volume). \\   
The measured rates at the end of the off-line data reduction process, without
correction for the detector livetime, are shown in Fig.~\ref{f:offline}, both
for single and multiple interaction events.  The count rate for single
interaction events is $\sim 0.7$~Hz while for multiple interaction events the
rate is $\sim$5~Hz.
\bfg[htb]
\centering
\psfig{file=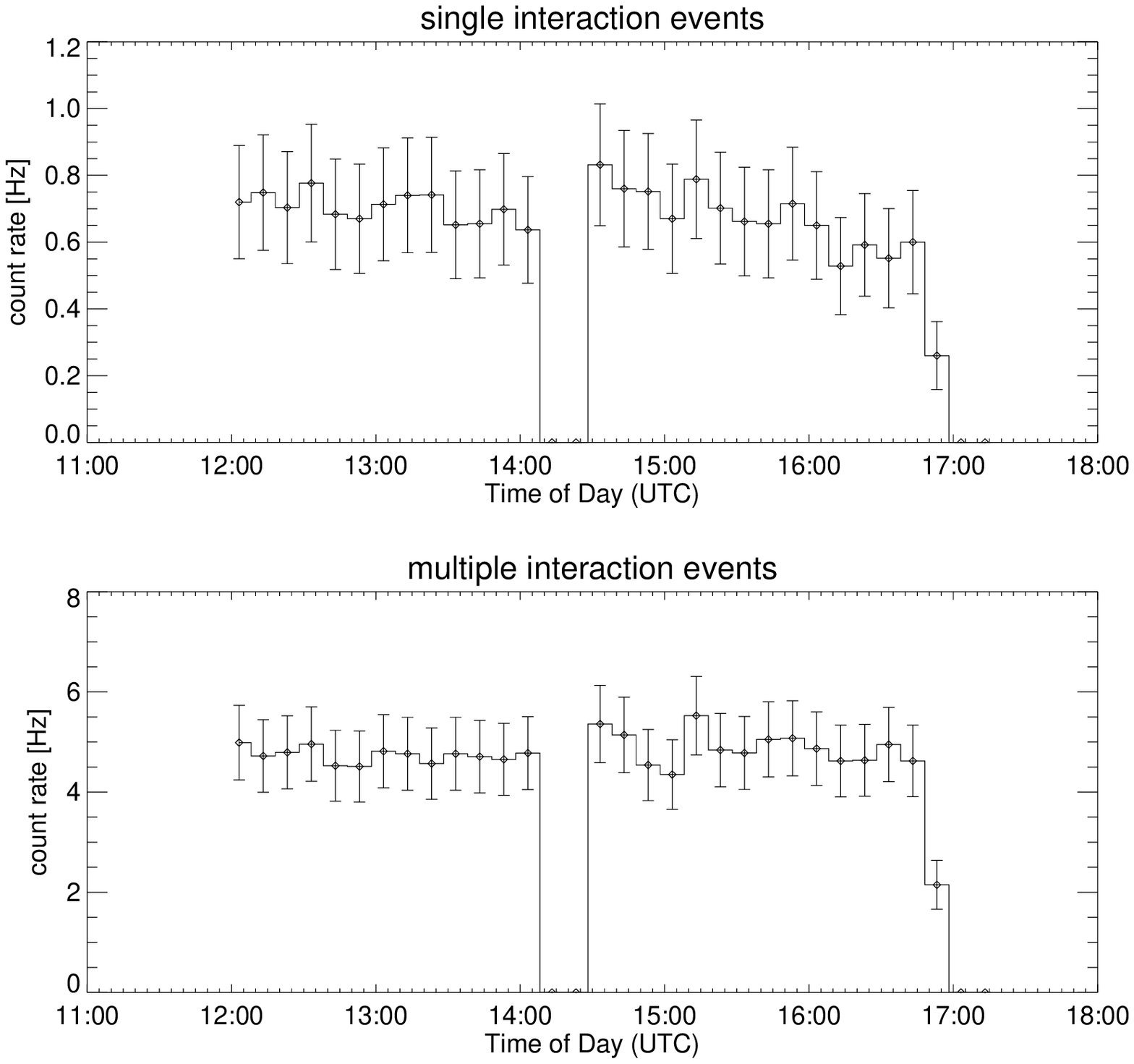,bbllx=79,bblly=285,bburx=554,bbury=732,width=.8\linewidth,clip=} 
\caption{\label{f:offline} \emph{Top:} measured rate for single interaction
events, \emph{bottom:} for multiple interaction events.} 
\efg
\section{LXeGRIT in-flight energy spectrum} \label{sec:infl_espec}

The energy spectrum of single site events recorded at balloon altitude in the
period 12:00 - 17:00~UT is shown in Fig.~\ref{f:intbkgd_2}, left panel. It is
largely dominated by low energy events, as expected (see Sec.~\ref{sec:MC} and
Sec.~\ref{sec:comp}). The lower energy threshold is $\sim$150~\keV, given by the
combined anode~/~wire signal-to-noise ratio. The high energy component is due to
single-scatter Compton events and to pair production events in which the two
511~\keV\ photons from positron annihilation both escape. The energy spectrum
for multiple interaction events is also shown in Fig.~\ref{f:intbkgd_2}, right
panel.
Compared to single site events, the spectrum is harder and the lower energy
threshold is higher ($\sim$300~\keV), for at least two interactions must be
above threshold. 
The clearly visible, especially in the multi-site events, $^{40}$K line is
attributed to natural radioactivity of the ceramic (MACOR, which contains about
10~\% potassium) used around the wire structure and field shaping rings. The
small amount of $^{40}$K (we estimated $\sim 10^{19}$ atoms of $^{40}$K) and the
clearness of the 1.465~\MeV\ line, with the expected resolution, indicate the
sensitivity of the detector. 
The \po\ line turned out to be a very useful in-flight
calibration tool.
The internal instrumental background, especially its low energy (below 500~\keV)
\g-ray component, constitutes a sizeable fraction of the measured
count rate. Repeated measurements in the laboratory have established the typical
background spectrum in the LXeTPC on the ground, and allowed us to identify internal background
sources. Subtraction of these contributions from the total
background at balloon altitude is straightforward. 
%
The \g-ray background as measured at ground level, with the payload on the
launch pad a few hours before the 2000 flight, is also shown in
Fig.~\ref{f:intbkgd_2}. The single site and multiple site energy spectra are
shown separately. The multiple site spectrum shows more clearly the \po\ line at
1.465~\MeV\ while the single site spectrum is largely dominated by a component
below 500~\keV. Due to the more complex assessment of the strongly suppressed
low-energy part of the single-site spectrum, we have not yet established how
much of the low energy continuum is due to internal background and how much is
due to local (environmental) \g-rays. The distinction is important since the one
due to ambient activity does not contribute to the background at balloon
altitude. In any case, this continuum is rather unimportant above several
hundred \keV.
 
In the study presented here the detector has been used as a spectrometer and
single site events are as useful as multiple site events. A localized internal
background source can be easily eliminated applying a fiducial volume cut. This
is the case of the localized, low energy events with z-position near  the
cathode (top of the detector - see Fig.~\ref{f:intbkgd_1}). These events are
likely due to $\alpha$-emission from the alumina substrate of the cathode plate 
\footnote{In a LXe detector simultaneously measuring ionization and
scintillation light like LXeGRIT does, $\alpha$ emission is identified by the
small charge~/~light yield ratio.} .
The same sharp (few mm) feature in the z-distribution has been detected both at
the ground level and in flight with comparable strength.
The fine granularity offers a very selective way to  reject this background in
the final data sample.  

\bfg[h]
\centering
\psfig{file=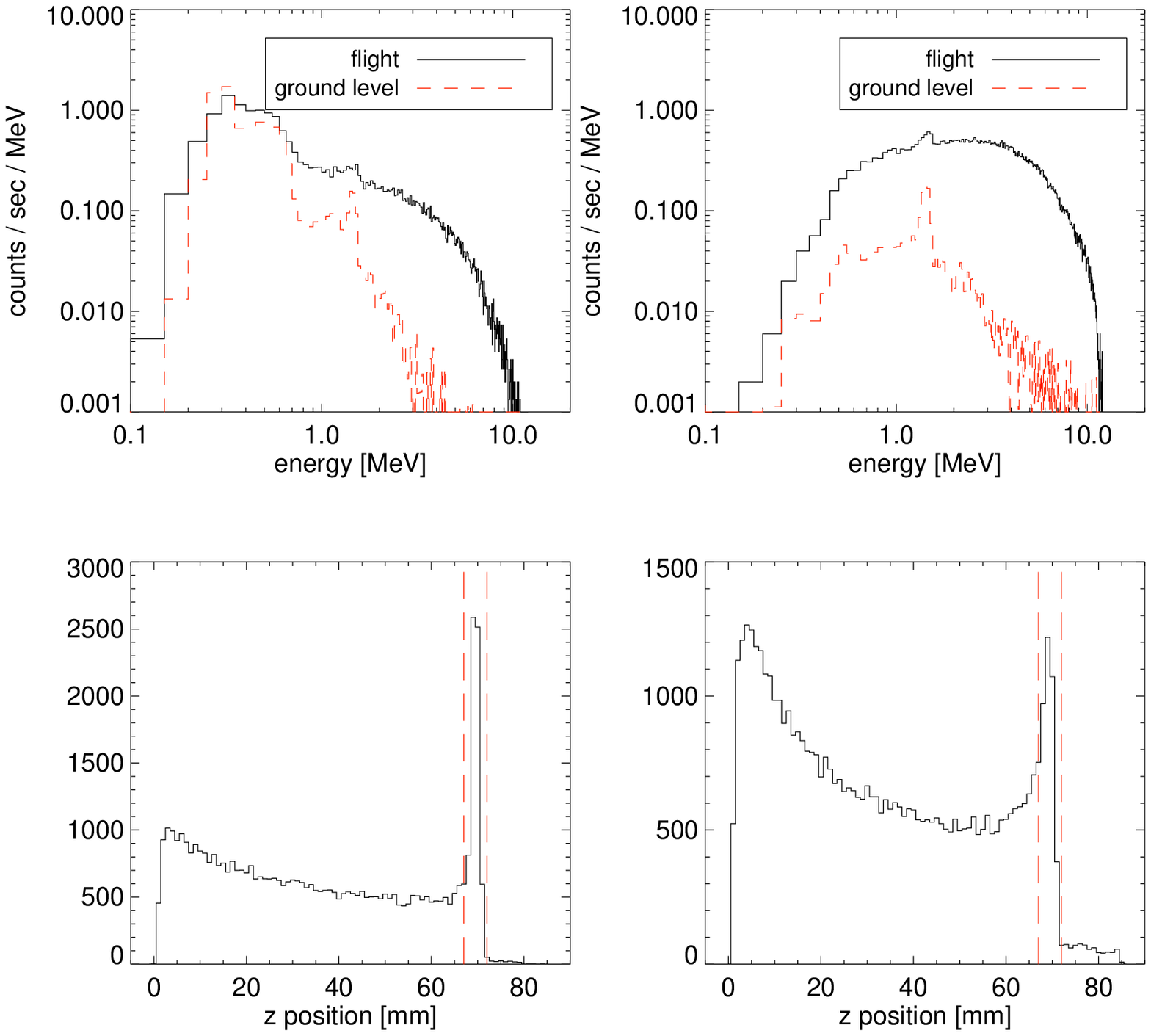,bbllx=61,bblly=515,bburx=544,bbury=718,width=\linewidth,clip=} 
\caption{\label{f:intbkgd_2} Energy spectra for flight data (\emph{continuous
line}) and for a background run at the ground level (\emph{dashed line}). {\it
Left:} single site events. \emph{Right:} multiple interaction events.}
\efg
\bfg[h]
\centering
\psfig{file=SPIE02.12.ps,bbllx=84,bblly=283,bburx=544,bbury=486,width=.9\linewidth,clip=} 
\caption{\label{f:intbkgd_1} \emph{Left:} Interaction depth (z coordinate) for
single site events at the ground level. The top of the TPC (cathode) is located
at z=70~mm. \emph{Right:} Interaction depth (z coordinate) for single site
events during flight. } 
\efg

\section{Monte Carlo simulation} \label{sec:MC}
\subsection{Mass Model}
We have studied the propagation of atmospheric and cosmic diffuse \g-rays
through a detailed model of the LXeGRIT instrument with a Monte Carlo simulation
based on GEANT 3.21. In the flight 2000 configuration, LXeGRIT was operated
without active shield and is thus an omni-directional detector in principle.
Moreover, in this study of single-site events, we use LXeGRIT as a mere
spectrometer without imaging information. It is therefore important to include
the bulk of passive materials of the entire instrument, including the gondola,
in the mass model, since these materials constitute significant mass for \g-ray
attenuation and scattering.
The total mass of the LXeGRIT instrument in flight configuration is
$\sim$1000~kg; the TPC itself, including the stainless steel cold vessel and
cryostat, is about 190~kg, of which 21~kg is liquid xenon. \\ 
The 20$\times$20$\times$7~cm$^3$ box shaped fiducial volume of the TPC is
enclosed in a 10 liter cylindrical vessel, filled with 7 liters of pure liquid
xenon. The remaining volume is filled with stainless steel blocks on three
sides and  a large ceramic HV feedthrough on the fourth side. The cold
vessel is enclosed in a vacuum cryostat to provide thermal insulation. Both the
vessel and the cryostat are made out of stainless steel and the walls are 3~mm
thick for both vessels. The cryostat thickness is 7~mm on top and 20~mm on
bottom; the vessel is 10~mm both on top and on bottom. On top of the TPC there
is a ``window'', i.e. both the cryostat and the vessel have been thinned (2~mm
and 5~mm in the z direction, respectively) over a circle of diameter
$\sim$20~cm.  
A layer of 5~mm of liquid xenon above the cathode and a layer of 3~cm below the
anodes have also been taken into account. 
Passive materials below the TPC cryostat, i.e. the structure of the gondola,
electronics boxes, LN$_2$ dewar, battery stack etc. have been modeled as three
aluminum plates of proper diameter and thickness located at different
z-positions. The simulated instrument is 155~cm high (z direction) with a
maximum diameter of 183~cm in the x-y plane (diameter of the bottom aluminum
plate).  

\subsection{Input Spectra}
\g-rays are started satisfying the spectral and geometrical properties of the
specific source under study. For the case of interest in this paper, cosmic
diffuse (CDG) and atmospheric (ATM) \g-rays were propagated through the LXeGRIT
mass model.
For the CDG we use the flux given by
Sch\"onfelder~et~al.~\cite{VSchonfelder.1980.CDG}~,
i.e. an isotropic flux over the zenith angles from 0\deg\ to 90\deg\ following
the power law 0.011~E$^{-2.3}$~cm$^{-2}$sec$^{-1}$sterad$^{-1}$\MeV$^{-1}$. 
\footnote{More recent work has established the CDG flux to be 5--10 times
smaller above 1~\MeV, e.g. Kappadath et al.\cite{SCKappadath.1996.CDG}~. At
these energies, however, the LXeGRIT background is fully dominated by
atmospheric background in the current simulation already.}
For the ATM, we follow the model given by Costa~et~al.~\cite{ECosta.1984.ATM},
for an atmospheric depth of 3~g~cm$^{-2}$.  The fluxes, in units of
cm$^{-2}$sec$^{-1}$\MeV$^{-1}$, are extrapolated in the energy range
100~\keV~-~15~\MeV\ with four angular bins for the following zenith angles:
\begin{itemize}
\item 0.03~E$^{-1.61}$ (0\deg\ - 45\deg)
\item 0.20~E$^{-1.48}$ (45\deg\ - 90\deg)
\item 0.43~E$^{-1.34}$ (90\deg\ - 135\deg)
\item 0.23~E$^{-1.51}$ (135\deg\ - 180\deg)
\end{itemize}
It is worthwhile stressing that the ATM flux is poorly known and suffers large
uncertainties. 

Each photon is propagated through the payload mass and, if the requirement of at
least one interaction in the TPC fiducial volume is satisfied, energy deposit,
location and interaction mechanism are recorded for each photon interaction.
Photoabsorption is dominant at energies below 250~\keV, while Compton scattering
dominates up to $\sim 6$~\MeV, so that in this energy band \g-rays are mainly
detected through multiple interactions. Above 6~\MeV, pair production becomes
increasingly important, together with secondary bremsstrahlung photons.

\subsection{LXeGRIT Response Function to ATM+CDG} \label{sec:resp}
The task of the Monte Carlo program is to follow the propagation of the \g-rays
through the various passive and active materials. For comparison with data, the
detector response function has to be accurately modeled and applied to the
physical interaction locations and energy deposits resulting from the basic MC
simulation. Such factors include energy thresholds, various efficiencies, as
well as realistic spatial and energy resolution. More specifically:
\begin{enumerate}
\item Light trigger efficiency: implemented on an event-by-event basis, using a
      look-up table as described in Sec.~\ref{sec:trigger}.
\item Efficiency of on-line selections at the second level trigger (MINMAX): the
      efficiency (or equivalently the rejection power) is routinely measured for
      each experiment and applied, accounting for its energy dependence (see
      Sec.~\ref{sec:trigger}).
\item A minimum energy threshold of 150~\keV\ for detection of each interaction. 
\item A minimum spatial separation of approximately of 5~mm on at least one
      coordinate between each pair of interactions in order to consider two
      interaction locations mutually resolved. If this condition is not
      fulfilled the two interactions are clustered and considered as one single
      interaction.
\item Energy resolution: 8.8~$\%$ (FWHM at 1~\MeV) scaling as $1/\sqrt{E}$ plus
      a noise term of 55~\keV\ (FWHM) added in quadrature.
\item DAQ livetime fraction.
\item Efficiency of off-line selections.
\end{enumerate}

Given the limited DAQ speed, the trigger efficiency during the 2000 flight was
reduced for low energy events. The DAQ livetime fraction is easily and precisely
accounted for, since it is measured by the DAQ processor itself.  Given its
complicated dependence on both position and energy, the light trigger efficiency
is applied on an event-by-event basis (see Sec.~\ref{sec:trigger}). The MINMAX
criterion has been parameterized as a function of energy only, assuming that
(statistically) Monte Carlo events reproduce the topologies of real events (this
is actually the case for calibration sources). In this way we do not test the
criterion on each event but simply correct the energy spectrum with a function
such as the one shown in Fig.~\ref{f:minmax}. Anode saturation is not considered
in the simulation since it impacts only charged particles. Comparison of Monte
Carlo data to calibration sources in the energy range 0.5~-~4.2~\MeV\ has shown
an excellent agreement.

\bfg[ht]
\centering
\psfig{file=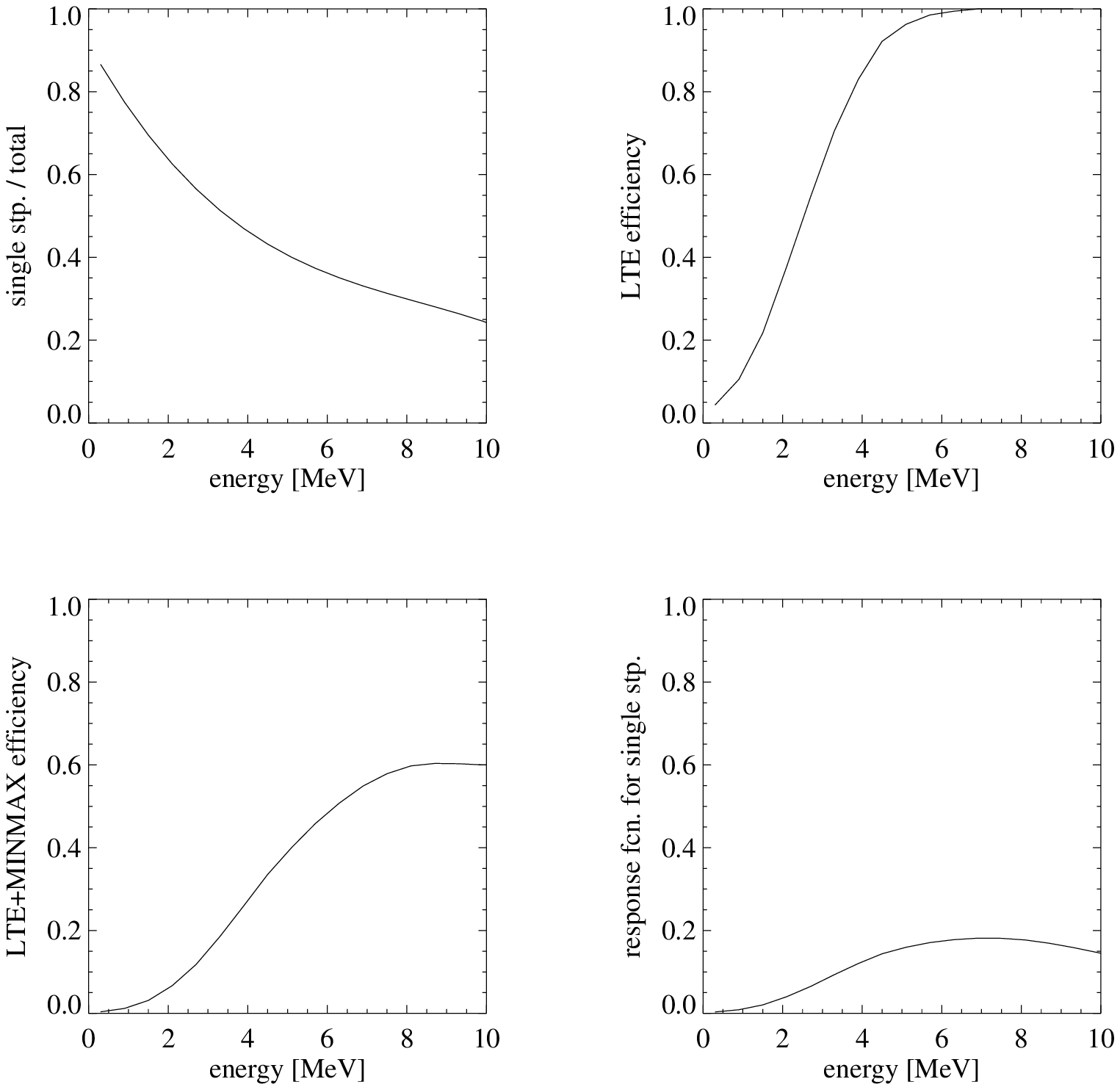,bbllx=91,bblly=290,bburx=529,bbury=717,width=.8\linewidth,clip=} 
\caption{\label{f:response} \emph{Top left:} selecting single site events, {\it
top right:} light trigger on single site events, \emph{bottom left:} light
trigger and MINMAX on single site events and, eventually, \emph{bottom right:}
single site selection, light trigger and MINMAX, i.e. the response function of
LXeGRIT to ATM+CDG once the \g-rays have been transported through the instrument
mass model. See text for more explanations.}
\efg

For this paper we have not attempted to carefully model all multiplicities
individually, but rather restrict ourselves to the case of single-site
events. The more extensive work to properly account for all multiplicities is
currently underway. Selecting single site events is part of the off-line
selections and is well reproduced in the Monte Carlo simulation, at least in the
energy range 0.5~-~4.2~\MeV, where a comparison to calibration data is
available. Fiducial volume selections are also easily accounted for in the
simulation. Event quality selections (e.g. rejection of noisy events) are not
considered in the present simulation. Several efficiencies entering the
instrumental response function are summarized in Fig.~\ref{f:response}.  It is
clear that, by design, the overall efficiency for single-site events is very
low, with a steep energy dependence.

\section{Comparison to Expectations} \label{sec:comp}

The expected background rate at balloon altitude can be split into two main
parts: \emph{a.} the total background rate, including charged particles, neutron
induced background etc. and \emph{b.} the ATM and CDG \g-ray background rate. 
The total background rate, with the 2000 flight settings, is basically given
by the PMT-OR trigger rates in Sec.~\ref{sec:flight}. To demonstrate the
capability of LXeGRIT to measure cosmic \g-rays it is important to measure the
ATM and CDG \g-ray contribution to the total background rate. A comparison to
previous experiments and models is also relevant and is discussed in this
section.    
We subdivide the total background in four categories:
\begin{itemize}
\item [1. ] charged particles
\item [2. ] neutron induced background
\item [3. ] instrumental background
\item [4. ] ATM+CDG
\end{itemize}

\subsection{Charged Particles (Primary and Secondary Cosmic Rays)}
\label{sec:chargedp} 

For a detector the size of LXeGRIT, the expected rate due to charged particles,
both primary and secondary cosmic rays, is a few hundred Hz. A more precise
estimate of the expected charged particle rate is beyond the scope of this paper
and requires a model of primary and secondary cosmic ray fluxes to be propagated
through the LXeGRIT mass model. Passive materials surrounding the TPC can
attenuate charged particles up to hundred \MeV, and cosmic rays may pass the
sensitive volume only peripherally, depositing correspondingly low amounts of
energy. In any case, we reject a large fraction of this component at the second
level trigger. The MINMAX criterion provides a fairly high rejection power;
events saturating the anode, i.e. energy deposits (order of) 10~\MeV\ or larger
are also rejected on-line. From a visual scan of data recorded without any
selections applied at the second level trigger, we can infer that charge
particles make about 2~/~3 of the total in-flight rate or about 400~Hz.  Events
not rejected on-line are rejected with an efficiency very close to one applying
offline selections on the anode waveform and on the fiducial volume (see
Sec.~\ref{sec:offline}).  To conclude, when measuring the \g-ray background, the
contamination due to charged particles is reduced to a \emph{truly negligible
level}.

\subsection{Neutrons}

Due to its complexity, the neutron induced background would deserve a
detailed and lengthy discussion on its own. A Monte Carlo simulation of the
expected rate in LXeGRIT due to the atmospheric neutron flux, based on the
GEANT3.21 / GCALOR~\cite{GCALOR} package and accounting for the induced \g-ray
emission - to be described elsewhere in greater detail - indicate a marginal
contribution to the overall rate. \\
From the data themselves, the lack of lines pointing to neutron activation
indicates that the neutron induced background is of little importance in
explaining the total background rate and, more specifically, the \g-ray
background at balloon altitude. This finding is not surprising, given the well
known radiation hardness of Xe compared to other scintillators or Ge, as found
in laboratory measurements, i.e. monoenergetic irradiation at accelerator
facilities, and in deep space (see Kirsanov~et~al.\cite{MAKirsanov:93} and
Ulin~et~al.\cite{SEUlin:98}).    

\subsection{Instrumental Background}

The instrumental background has been partially discussed in
Sec.~\ref{sec:infl_espec}. The spallation local production background at balloon
altitude is however not known. A preliminary calculation indicates that it is
rather negligible compared to CDG and ATM fluxes, and we do not measure any
appreciable excess in the count rate. A more detailed calculation is under way.

\subsection{ATM+CDG} \label{sec:atm+cdg}

Using the Monte Carlo simulation described in Sec.~\ref{sec:MC} to transport the
\g-rays and assuming an energy threshold of 150~\keV, the expected ATM and CDG
interaction rate in the XeTPC is $\sim$100~Hz. The atmospheric background is
the dominant component, while the cosmic diffuse contributes $\sim$10~$\%$
of the total interaction rate at energies below 500~\keV\ and becomes
much less important at higher energies. Up to this point only the geometry
and the mass model of the instrument are taken into account. 
These numbers are very far from the actual measured rate (see
Sec.~\ref{sec:offline}), since we do not yet include the instrument response
function (Sec.~\ref{sec:resp}).
Therefore, in order to get to the final comparison we still have to correct the
Monte Carlo prediction with the detector response function, which is applied
event-by-event (the one for the ATM+CGD is summarized in Sec.~\ref{sec:resp})
The outcome of this analysis is shown in Fig.~\ref{f:comp}. 
The Monte Carlo expectation still overestimates the measured rate and an overall
factor of 2 has been introduced to account for efficiency losses due to off-line
rejection of noisy events and event quality selections. \\ 
The agreement is fairly good even if a certain discrepancy is present for
energies lower than 1.5~\MeV. This is mainly due to the fact that, at these
energies, the intrinsic internal background, which is not included in the Monte
Carlo simulation, plays some role (e.g. the prominent \po\ line).
The discrepancy is well within uncertainties in our model of the instrument and
in the input fluxes.

\bfg[h]
\centering
\psfig{file=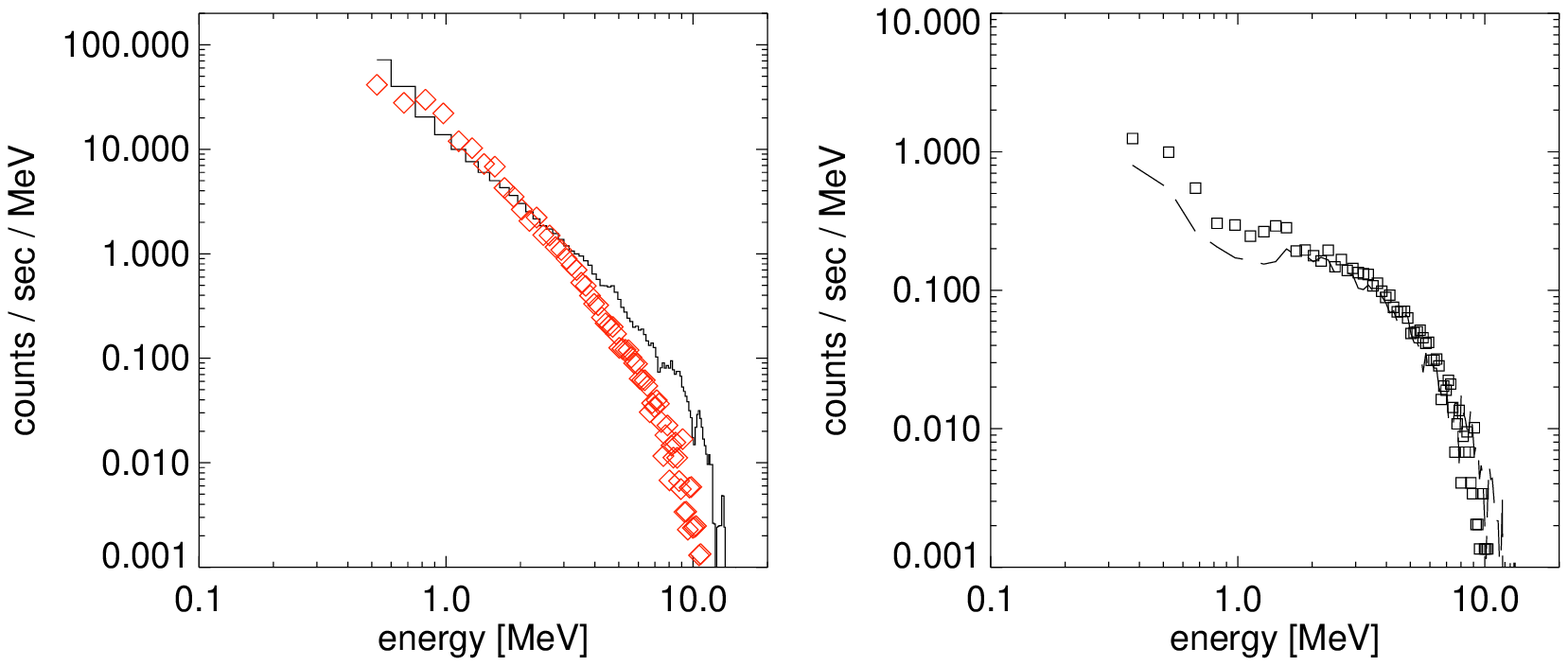,bbllx=301,bblly=515,bburx=533,bbury=716,width=.5\linewidth,clip=} 
\caption{\label{f:comp} \emph{Open squares} - Experimental data. 
\emph{Dashed line} - Monte Carlo prediction. See text for explanation.}  
\efg

\section{Conclusions}
In this paper we have presented a measurement of the \g-ray background rate in
the LXeGRIT Compton telescope from a sample of data acquired at 3.2~g~cm$^{-2}$
atmospheric depth during the 2000 balloon flight from Ft.~Sumner, NM. The
measured trigger rate at float altitude was about 600~Hz at the first level
trigger. Of this about 400~Hz were charged particles, most of which were
rejected online. Of the remaining rate, we studied the single-site \g-ray
events and compared them with the rate expected from Monte Carlo simulations of
the known atmospheric and cosmic diffuse background. We find a good
agreement between data and expectation.  This result is very encouraging for the
LXeTPC detector technology for the field of \MeV\ \g-ray astronomy.

\acknowledgments  

We would like to thank Chun Zhang for his
contribution in the preparation of the balloon flight in year 2000. \\
This work was supported by NASA under grant NAG5-5108.

\small
\bibliography{bbl}  
\bibliographystyle{spiebib}   


\end{document}